\documentclass[preprint,aps]{revtex4}
\usepackage{mathrsfs}

\usepackage{graphicx}

\begin{document}

\title{Evidence of Topological Surface State in Three-Dimensional Dirac Semimetal Cd$_3$As$_2$}

\author{Hemian Yi$^{1}$, Zhijun Wang$^{1}$, Chaoyu Chen$^{1}$, Youguo Shi$^{1}$, Ya Feng$^{1}$, Aiji Liang$^{1}$, Zhuojin Xie$^{1}$, Shaolong He$^{1}$, Junfeng He$^{1}$, Yingying Peng$^{1}$, Xu Liu$^{1}$, Yan Liu$^{1}$, Lin Zhao$^{1}$, Guodong Liu$^{1}$, Xiaoli Dong$^{1}$, Jun Zhang$^{1}$, M. Nakatake$^{3}$, M. Arita$^{3}$, K. Shimada$^{3}$, H. Namatame$^{3}$, M. Taniguchi$^{3}$, Zuyan Xu$^{4}$, Chuangtian Chen$^{4}$, Xi Dai$^{1}$, Zhong Fang$^{1}$  and X. J. Zhou$^{1,2,*}$}

\affiliation{
\\$^{1}$Beijing National Laboratory for Condensed Matter Physics, Institute of Physics, Chinese Academy of Sciences, Beijing 100190, China
\\$^{2}$Collaborative Innovation Center of Quantum Matter, Beijing, China
\\$^{3}$Hiroshima Synchrotron Radiation Center, Hiroshima University, Hiroshima 739-8526, Japan
\\$^{4}$Technical Institute of Physics and Chemistry, Chinese Academy of Sciences, Beijing 100190, China.
}

\date{May 22, 2014}

%%\begin{abstract}

%%\begin{abstract}

\maketitle

{\bf
The three-dimensional topological semimetals represent a new quantum state of matter.  Distinct from the surface state in the topological insulators that exhibits linear dispersion in two-dimensional momentum plane, the three-dimensional semimetals host bulk band dispersions linearly along all directions, forming discrete Dirac cones in three-dimensional momentum space. In addition to the  gapless points (Weyl/Dirac nodes) in the bulk, the three-dimensional Weyl/Dirac semimetals are also characterized by ``topologically protected" surface state with Fermi arcs on their specific surface.  The Weyl/Dirac semimetals have attracted much attention recently they provide a venue not only to explore unique quantum phenomena  but also to show potential applications. While Cd$_3$As$_2$ is proposed to be a viable candidate of a Dirac semimetal, more experimental evidence and theoretical investigation are necessary to pin down its nature. In particular, the topological surface state, the hallmark of the three-dimensional semimetal, has not been observed in Cd$_3$As$_2$. Here we report the electronic structure of Cd$_3$As$_2$ investigated by angle-resolved photoemission measurements on the (112) crystal surface and detailed band structure calculations.  The measured Fermi surface and band structure show a good agreement with the band structure calculations with two bulk Dirac-like bands approaching the Fermi level and forming Dirac points near the Brillouin zone center.  Moreover, the topological surface state with a linear dispersion approaching the Fermi level is identified for the first time. These results provide strong experimental evidence on the nature of topologically non-trivial three-dimensional Dirac cones in Cd$_{3}$As$_{2}$.
}

%%%%\pacs{73.20.-r, 73.20.At, 71.18.+y}
% PACS, the Physics and Astronomy Classification Scheme.
\maketitle

The Dirac materials, such as graphene and topological insulators, exhibit novel quantum phenomena and properties with their low energy electron behaviors described by the relativistic Dirac equation\cite{Vafek,Volovik}.  The massless Dirac fermions on the surface of graphene\cite{Geim07, Castro, Geim09} and the three-dimensional topological insulators\cite{Hasan,XLQi} show cone-shaped band dispersion with the Dirac points lying at discrete points in two-dimensional (2D) momentum plane.  Recently, much attention has been attracted to the Weyl/Dirac semimetals which represent a new topological state of quantum matter that is different from the topological insulators\cite{ZFang,GXu,XGWan,Balents11,Balents11w,Balents12,SMYoung,ZJWang1,Turner,Shuichi}. These so-called three-dimensional (3D) Weyl/Dirac semimetals have inverted band structures in which the bulk valence band and conduction band contact at discrete points (Dirac points) in three-dimensional momentum space.  They can be viewed as the 3D analogue of graphene and the surface state of the topological insulators that host Dirac band dispersions linearly along all directions\cite{SMYoung, ZJWang}. In addition to the  gapless points (Weyl/Dirac nodes) in the bulk that are ``topologically protected" by the behavior of the band structure,  the Weyl/Dirac semimetal is also characterized by ``topologically protected" surface state with Fermi arcs on the surface\cite{XGWan,GXu}.  The Weyl/Dirac semimetal is expected to show various new quantum phenomena and unusual properties like magnetic monopole in the bulk, giant diamagnetism, quantum spin Hall effect or quantum anomalous Hall effect in its quantum well structure\cite{ZFang,Koshino,ZJWang1,CXLiu,GXu}.  In the case with both the time reversal and inversion symmetries, one may expect a 3D Dirac semimetal state described as four-component Dirac fermions which can be viewed as two copies of distinct Weyl fermions\cite{Volovik1}. To get Weyl semimetal, either the time reversal symmetry or the inversion symmetry has to be broken.

%%Uniqueness of Cd3As2
While a number of materials and systems have been proposed for realizing the Weyl/Dirac semimetal, few has been materialized so far\cite{XGWan,GXu,SMYoung,ZJWang1,ZJWang,Sato,SYXu}.  Among the proposals, Cd$_{3}$As$_{2}$ has been paid a special attention for a couple of reasons\cite{ZJWang}. First, band structure calculations predict that Cd$_3$As$_2$ is a symmetry-protected topological semimetal, with a single pair of 3D Dirac points in the bulk and nontrivial Fermi arcs on the surface\cite{ZJWang}.  The pseudo-relativistic physics of dispersions results in 3D Dirac points that are protected by crystal symmetry and robust against perturbations. No gap can be opened by the spin-orbit coupling, in contrast to Dirac materials like AMnBi$_{2}$(A= Sr,Ca)\cite{Park,YFeng}. Second, Cd$_3$As$_2$ is chemically stable in air which makes it more convenient for property investigations and potential applications when compared to other candidates\cite{ZJWang1,YLChen1}. Third, similar to graphene, Cd$_{3}$As$_{2}$ has particularly high carrier mobility up to 15000 $cm^{2}V^{-1}s^{-1}$ at room temperature and 80000 $cm^{2}V^{-1}s^{-1}$ at 4 K\cite{Zdanowicz,Gerin}, which makes it ideal for revealing novel quantum phenomena from transport measurements as well as designing electronic devices. It provides a promising platform to realize quantum spin Hall effect and sizable linear quantum magnetoresistance even up to room temperature\cite{ZJWang}. While the indications on the 3D Dirac fermion nature of Cd$_3$As$_2$ are accumulating, with some angle-resolved photoemission results reported for the Cd$_3$As$_2$ (001) surface\cite{Borisenko,MNeupane} and (111) surface\cite{YLChenCdAs}, more experimental evidence and theoretical investigation are necessary to pin down the issue. In particular, the topological surface state, the hallmark of the 3D semimetal, has not been identified in Cd$_3$As$_2$.  If the 3D Dirac fermion nature of Cd$_3$As$_2$ can be fully proven, it will open an opportunity to expand further into a Weyl semimetal when the time-reversal or inversion symmetry is broken.

%%Experimental description
Angle-resolved photoemission spectroscopy (ARPES) is a powerful tool which can directly reveal the presence of the Dirac cone and its 3D character. Here we report our ARPES study and band structure calculations on the (112) crystal surface of Cd$_{3}$As$_{2}$.  Linear-dispersive energy bands are observed in the low energy region that are associated with the bulk Dirac nodes. Such linear-dispersive behavior of the bulk bands is further examined along the out-of-plane direction by varying the photon energy of the light source. In particular, we have identified for the first time the topological surface state in the gapless bulk Dirac nodes of Cd$_3$As$_2$. These results are consistent with the band structure calculations and provide strong indications of topologically non-trivial 3D Dirac cones in Cd$_{3}$As$_{2}$.

%%Fig. 1: Demonstration of (112) cleaving surfac of Cd3As2

Figure 1 shows the crystal structure, Brillouin zone and X-ray diffraction patterns of Ca$_3$As$_2$ (see Supplementary for experimental details). The crystal structure of Cd$_{3}$As$_{2}$ is body-centered tetragonal with an I41cd symmetry; it is related to the tetragonally-distorted anti-fluorite structure with 1/4 Cd site of ordered vacancy\cite{Steigmann}.  As seen from the small cubic sub-cell (bottom of Fig. 1a), the As ions are approximately close-packed and the Cd ions are tetrahedrally coordinated. The overall crystal structure of Cd$_3$As$_2$ (top of Fig. 1a) can be viewed as a 2$\times$2$\times$4 superstructure stacked from the small cubic sub-cell.  Fig. 1b shows the reduced Brillouin zone and the projected (112) surface Brillouin zone for the small unit cell of Cd$_{3}$As$_{2}$ that is used in our band structure calculations. Powder X-ray diffraction pattern (Fig. 1c) on the sample from grinding the grown Cd$_{3}$As$_{2}$ single crystals provides an accurate determination of the lattice parameters a=b= 12.6467{\AA} and c=25.4428{\AA}. Fig. 1d displays the XRD pattern of the Cd$_{3}$As$_{2}$ single crystal. The observed peaks can be indexed into (n n 2n) (with n being integers) reflections which indicate that the naturally cleaved surface is (112). It actually corresponds to the close-packed As planes in the small cubic sub-cell with a distance of 3.68 {\AA} between two adjacent planes.

%%Fig. 2: Fermi surface of Cd3As2 (112)
Figure 2 shows constant energy contours of the spectral weight distribution for the cleaved (112) surface of Cd$_{3}$As$_{2}$ crystal over a large momentum space (Fig. 2a), and their comparison with band structure calculations (Fig. 2b). Here the plotted hexagonal Brillouin zone is the projected (112) surface Brillouin zone (Fig. 1b). The spectral weight repetition over three zones reaffirms that the measured crystal surface is (112). Right at the Fermi level, the measured spectral weight is point-like that concentrates at the Brilllouin zone centers ($\Gamma$ points) (Fig. 2a1). With increasing binding energy, the spectral weight distribution gets expanded and at high binding energies like 450 meV (Fig. 2a4) and 600 meV (Fig. 2a5), two fine features, the inner part and outer part, can be discerned as marked by blue dashed lines. The inner part appears as a small circle. The outer part changes from a point-like Fermi surface to a ring-like circle and eventually to a hexagon-like shape at high binding energy. From the band structure calculations, there are only Fermi points at the Fermi level (Fig. 2b1). With increasing binding energy, the point gradually becomes hexagon-shaped at 150 meV, and warped hexagon-shape at high binding energies. The spectral weight evolution with increasing binding energy is reminiscent with many materials with Dirac cones such as graphene and topological insulators. The hexagon-shape indicates that the related Dirac cone is anisotropic, a situation similar to that found in Bi$_2$Te$_3$\cite{YLChen09,CYChen}. Although the calculated fine-details of the electronic structure are not fully resolved in our measurements, our results agree well with the band structure calculations. Overall, we find that the periodicity of the ARPES signal can be well described by using the small sub-cell (Fig. 1a); the effect of the (2$\times$2$\times$4) super-cell that is supposed to cause band folding is rather weak.

%%Fig. 3: Band structure of  Cd3As2£¨112£©: consistence with band structure calculations and identificaition of various bands
Figure 3 shows band structures of Cd$_3$As$_2$ (112) surface measured along two high-symmetry directions. There is a good agreement between the measured band structures (Figs. 3b and 3c) and the band structure calculations (Fig. 3d). Along the $\Gamma$-K cut, from the calculation, there are four main bulk bands, labeled as BV1, BV2, BV3 and BV4, within the energy range shown (Fig. 3d).  Along the $\Gamma$-M cut, there are three bulk bands, labeled as BV5, BV6 and BV7(Fig. 3d). As seen from the measured band structures (Figs. 3b and 3c) along $\Gamma$-K (cut B, green line in Fig. 3a) and $\Gamma$-M (cut A, red line in Fig. 3a), all these seven bulk bands from the calculations can find good correspondence. It is clear that, within the similar energy range, we have resolved more bands than reported before on the (112) surface of Cd$_3$As$_2$\cite{YLChenCdAs}. The good agreement between the measured and calculated band width also indicates that the band renormalization effect in Cd$_3$As$_2$ is weak.  Furthermore, there is a surface band from the calculated results (labeled SV1 in Fig. 3d) that appears near the zone center ($\Gamma$ point). The surface state only appears in surface band calculation when taking into account the spin-orbit coupling as well as the inverted bulk band structure near the Fermi level.  As seen from Figs. 3b and 3c, there is a band in the measured data labeled as SV1 that is in good agreement with the calculated surface band.  We note that there is some band intensity asymmetry along the equivalent momentum cut locations. For example, as seen in Fig. 3b, along $\Gamma_1$-M, three bands are clearly seen but along the equivalent $\Gamma_{21}$-M, the BV5 band gets stronger and the BV6 band becomes invisible. For the SV1 band, it appears clearly only on one side of $\Gamma$-M (Fig. 3b) and $\Gamma$-K (Fig. 3c) measurements. These intensity variations can possibly be attributed to the photoemission matrix element effect.  From the measurements, there are three bands, two bulk bands(BV2 and BV5) and one surface band (SV1), that are Dirac-cone-like approaching the Fermi level.  The good agreement between the measurements and band structure calculations provide an unambiguous identification of each band in Cd$_3$As$_2$ that lays a foundation for our further study below.

%%Identifications of surface state
The nature of the surface band SV1 can be further examined by varying the photon energy during the ARPES measurements to detect band structures at different K$_z$ planes. As seen in Fig. 4a, at different photon energies, the SV1 band (marked as dashed blue lines) shows little change within the experimental uncertainty. This again provides an additional strong evidence on the nature of the SV1 surface band. We also find that this surface state is rather robust as it does not show obvious change during long time ARPES measurement even at a high temperature ($\sim$300 K).  Interestingly, the SV1 surface band lies within the gap region formed by the BV2 and BV3 bands along the $\Gamma$-K and BV5 and BV6 bands along $\Gamma$-M.  This is distinct from the topological surface state in 3D topological insulators such as Bi$_2$Se$_3$\cite{YXia} and Bi$_2$Te$_3$\cite{Hasan}. This surface state possesses a Fermi velocity as high as 7.5eV which has nearly the same value along G-K and G-M directions, suggesting an isotropic nature of the SV1 surface state. These results are consistent with the existence of a two-dimensional Dirac cone derived from the SV1 surface band.  The revelation of the topological surface state is in agreement with the expectation from theoretical calculations on the non-trivial bulk band topology of Cd$_{3}$As$_{2}$\cite{ZJWang}.

The existence of possible 3D Dirac fermion in  Cd$_{3}$As$_{2}$, which is supposed to disperse linearly along all three directions in momentum space, can be checked by the two-dimensional energy dispersions in the K$_x$-K$_y$ basal plane by varying K$_y$ values (Fig. 4b) and the band dispersion along K$_z$ by varying the photon energy (Fig. 4a). As seen from Fig. 4b, although the signal is weak, there is an overall trend that, with the momentum cut moving along K$_y$ within the K$_x$-K$_y$ basal plane (cuts B1 to B5 in Fig. 4c), the cross point of the BV1 bulk band approaches the Fermi level at K$_y$=0 and sinks below the Fermi level when the momentum cut moves away from the $\Gamma$-K direction. The dispersion can be fitted with a linear line at the Brillouin zone center (cut 3).  Away from the Dirac point, the dispersions can be fitted with hyperbola lineshape, similar to the Dirac cone formed on the surface of topological insulators\cite{Hasan,XLQi}.  The measured momentum dependence is consistent with the calculated result (Fig. 4d).  These results are consistent with the presence of two-dimensional Dirac fermion from the BV1 band in the K$_x$-K$_y$ plane.  On the other hand, as seen from Fig. 4a, the BV5 bulk band changes its position sensitively with the photon energy. It reaches the Fermi level near photon energies of 21.2 eV (Fig. 4a1) and 97 eV (Fig. 4a4) where Dirac-like linear bands are formed. For other photon energies, the cross point of the band appears below the Fermi level.  Although we do not know the exact K$_z$ values corresponding to these photon energies, the observation of the Dirac-like bands at some discrete photon energies (K$_z$s) is consistent with the theoretical expectations that the conduction and valence bands touch only at discrete points in the momentum space.  As the bulk Cd$_3$As$_2$ has I41cd symmetry identified for the X-ray diffraction patterns, in principle, the Dirac points are along the [001] direction\cite{ZJWang} due to the C4 symmetry. On the (112) surface, however, the Dirac points is located along the [112] direction. This is possible because APPES is mostly sensitive to the surface layers that has C3 symmetry as observed in the Fermi surface of this surface in Fig. 2. It is quite unique for Cd$_3$As$_2$ (112) surface that the bulk Dirac cone, as well as the surface Dirac cone, lies close to the Fermi level, indicating its electric neutrality that is ideal for investigating the intrinsic non-trivial properties.

In summary, we report angle-resolved photoemission measurements and band structure calculations on (112) surface of Cd$_{3}$As$_{2}$. We find that the overall measured Fermi surface and band structure show a good agreement with the band structure calculations. We have revealed the existence of two-dimensional Dirac fermions from topological surface state at the (112) crystal surface of Cd$_{3}$As$_{2}$. We also present experimental results of 3D Dirac fermions in Cd$_{3}$As$_{2}$ by observing the linear-dispersive bulk valence band along all the three directions in the momentum space. These results are consistent with the band structure calculations which indicate that Cd$_{3}$As$_{2}$ is a three-dimensional Dirac semimetal. Further high resolution ARPES work needs to be done to search for the arc-type Fermi surface as expected from theory\cite{ZJWang}.

$^{*}$Correspondence and requests for materials should be addressed to X.J.Z. (XJZhou@aphy.iphy.ac.cn).\\

\vspace{3mm}

\noindent {\bf Acknowledgement}\\
%%\begin{acknowledgments}
We thank Desheng Wu for preparing single crystal and discussion. This work is supported by the National Natural Science Foundation of China (91021006, 10974239, 11174346 and 11274367) and the Ministry of Science and Technology of China (2011CB921703, 2013CB921700 and 2013CB921904).  The synchrotron radiation experiments were performed with the approval of HSRC (Proposals No. 12-B-47 and No. 13-B-16).
%%\end{acknowledgments}

%\noindent {\bf Supplementary Information}  is linked to the online version of the paper.

\vspace{3mm}

%%%%\vspace{3mm}

\noindent {\bf Author Contributions}\\
 H.M.Y. and X.J.Z. proposed and designed the research.  Z.J.W, X.D. and Z.F. carried out band structure calculations. Y.G.S. grew the single crystal Cd$_3$As$_2$ samples. H.M.Y.,C.Y.C.,Y.F., A.J.L., Z.J.X., S.L.H., Y.Y.P., X.L., Y.L., L.Z., G.D.L., X.L.D., J.Z., Z.Y.X., C.T.C. and X.J.Z. contributed to the development and maintenance of Laser-ARPES system. M.N., M.A., K.S., H.N.and M.T. contributed in synchrotron radiation ARPES measurements. H.M.Y., C.Y.C., Y.F., A.J.L.,Z.J.X. and S.L.H. carried out the experiment.  X.J.Z.and H.M.Y. wrote the paper. All authors participated in discussion and comment on the paper.

\vspace{3mm}

\noindent{\bf Competing financial interests:} The authors declare no competing financial interests.

\newpage

\newpage

\begin{figure*}
\centering
\includegraphics[width=1.0\columnwidth]{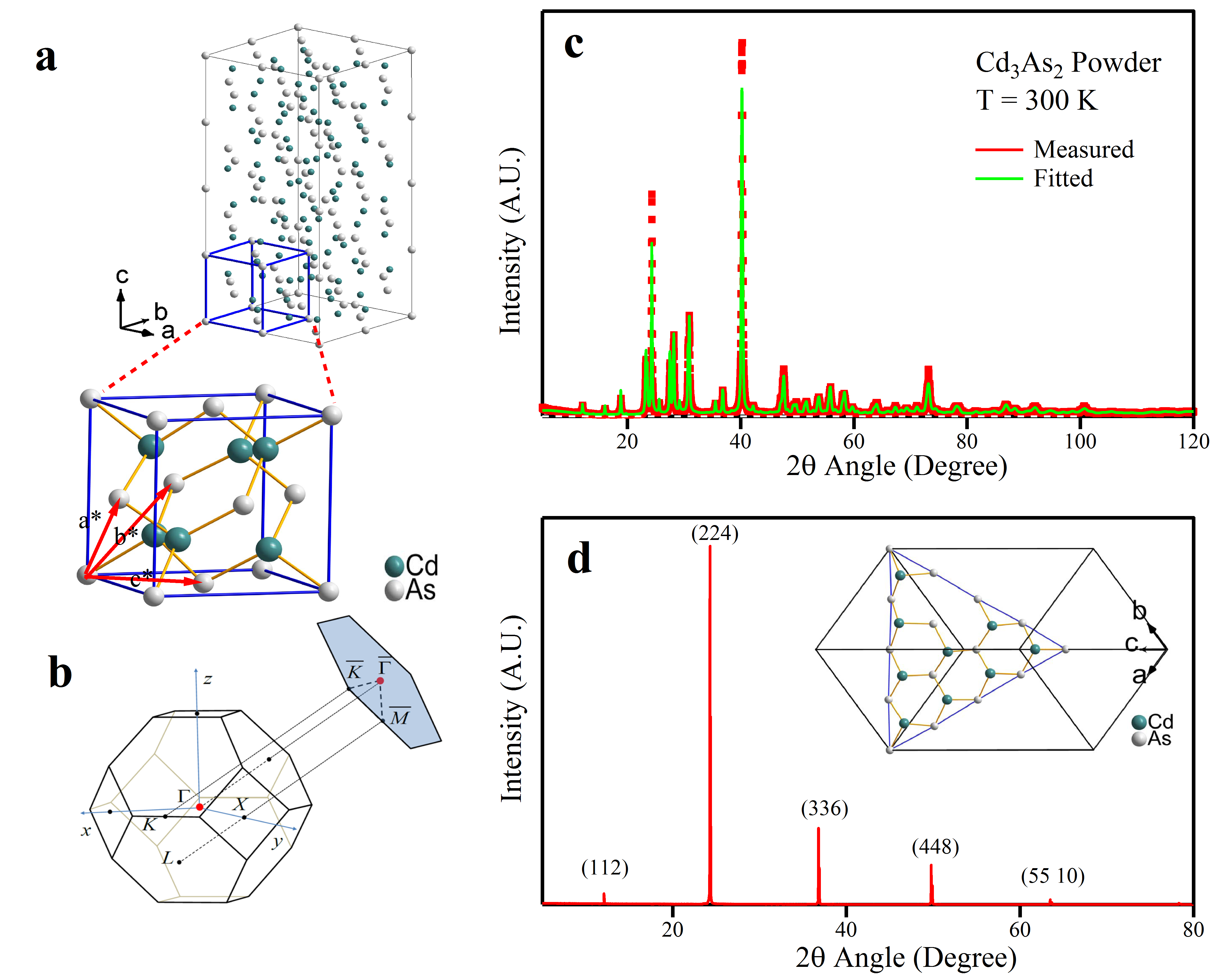}
\caption{\label{fig1} Crystal structure, Brillouin zone and the cleaved surface of Cd$_3$As$_2$. (a). Crystal structure of Cd$_{3}$As$_{2}$ with a body-centered tetragonal structure (Space group: I41cd). Such a unit cell can be viewed as a re-constructed 2$\times$2$\times$4 superstructure from a small sub-cell expanded in the bottom-left. The As ions form a face-centered cubic while the Cd ions fill 3/4 of the 8 tetragonal sites formed by the As ions. (b). The bulk Brillouin zone and projected (112) surface Brillouin zone used in the band structure calculations which corresponds to the smallest unit cell marked in the sub-cell as red arrows in (a).   (c). The X-ray diffraction pattern of Cd$_3$As$_2$ powder taken with Cu K$\alpha$ radiation ( $\lambda$= 1.5418 \AA). The measured spectrum (red squares) is compared with the calculated result (green line) from the structure refinement. It confirms the crystal structure with a space group I41cd; the lattice constants obtained are a=b=12.6467{\AA} and c=25.4428{\AA}. (d). X-ray diffraction pattern of a naturally cleaved surface of the Cd$_3$As$_2$ single crystal that indicates the (112) surface.}
\end{figure*}

\begin{figure*}
\centering
\includegraphics[width=1.0\columnwidth]{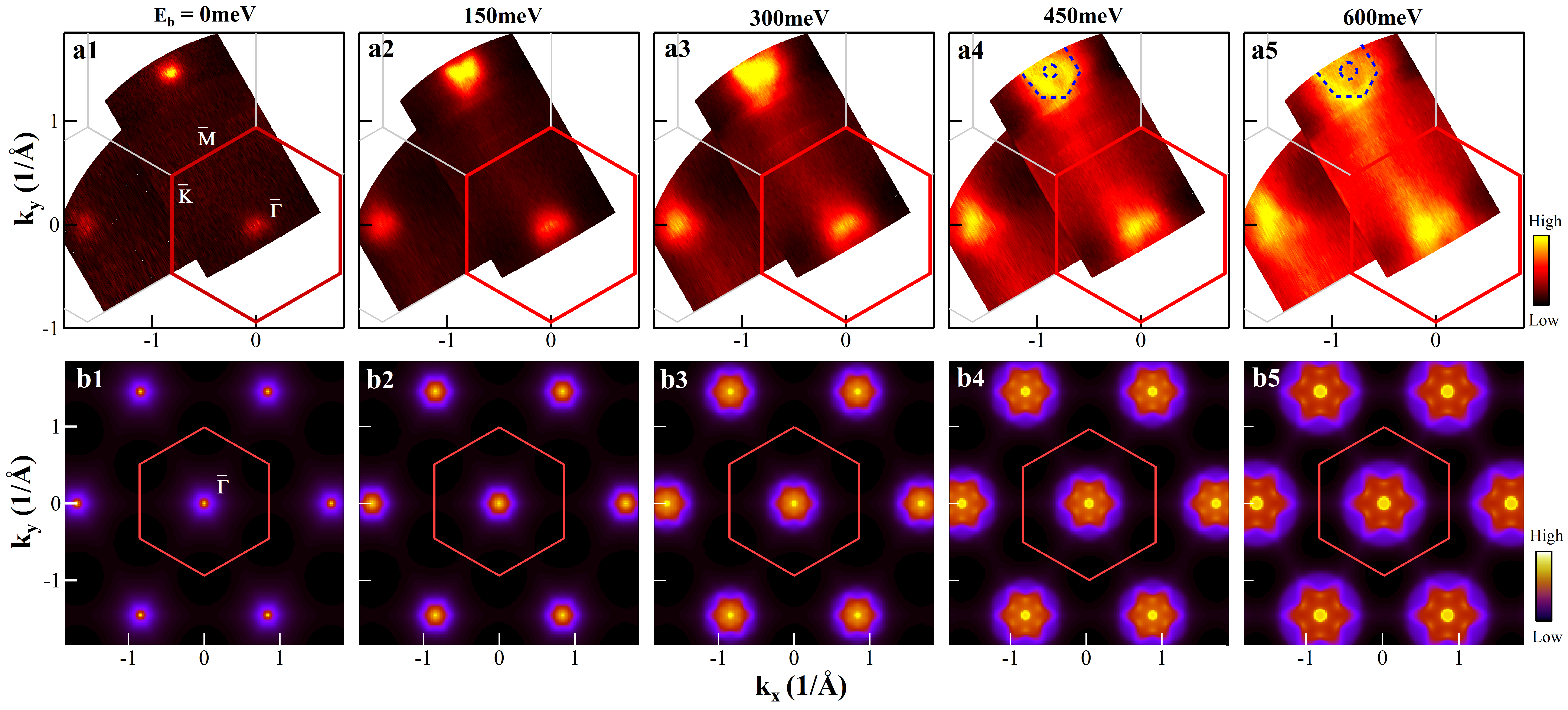}
\caption{\label{fig2}Fermi surface of the Cd$_3$As$_2$ (112) surface and its comparison with band structure calculations. (a). Constant energy contours of Cd$_{3}$As$_{2}$ measured using 21.2 eV light source at a temperature of 45 K.  The hexagons depicted in the images represent the (112) surface Brillouin zones with the first zone plotted as red. They are calculated with lattice constant a=4.45{\AA} for the (112) crystal surface. The measured momentum space covers three zone centers.  The small dashed hexagon and circle in (a4) and (a5) are guides to the eye.  (b). The constant energy contours of Cd$_{3}$As$_{2}$ from the band structure calculations. The small cell unit calculation takes into account the inverted band structure and modification of the band occupation according to the real crystal case. There are two small ring-like Fermi surface sheets around $\Gamma$. With increasing binding energy, the outer sheet becomes warped hexagon-shaped at a binding energy of E$_{b}$=600 meV while the inner one keeps its ring-like shape.
}
\end{figure*}

\begin{figure*}
\centering
\includegraphics[width=1.0\textwidth]{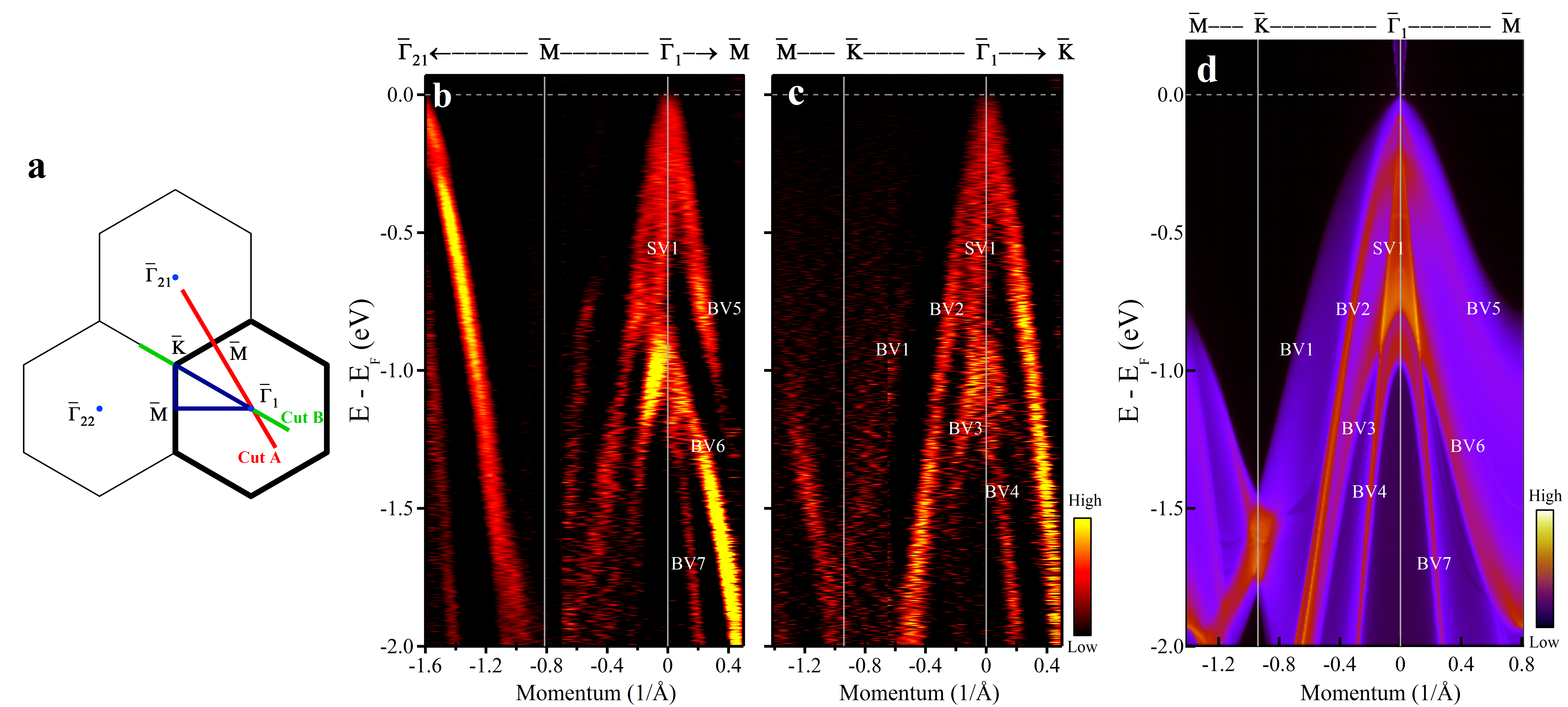}
\caption{\label{fig3} Band structures measured along high symmetry cuts for Cd$_3$As$_2$ (112) surface and their comparison with the band structure calculations.  (a). Location of two momentum cuts along $\Gamma$-M direction (cut A, red line) and $\Gamma$-K direction (cut B, green line).  (b). Band structure of Cd$_{3}$As$_{2}$ (112) surface measured along high symmetry M-$\Gamma$-M direction (cut A in (a)). (c).  Band structure measured along M-K-$\Gamma$-K direction (cut B in (a)).  To highlight the measured bands more clearly, (b) and (c) are second-derivative images with respect to the energy. (d).  Calculated band structure along M-K-$\Gamma$-M high symmetry lines (blue lines in (a)). Several bulk valence bands (BV) and a surface band (SV) are labeled.
}
\end{figure*}

\begin{figure*}
\centering
\includegraphics[width=1.0\textwidth]{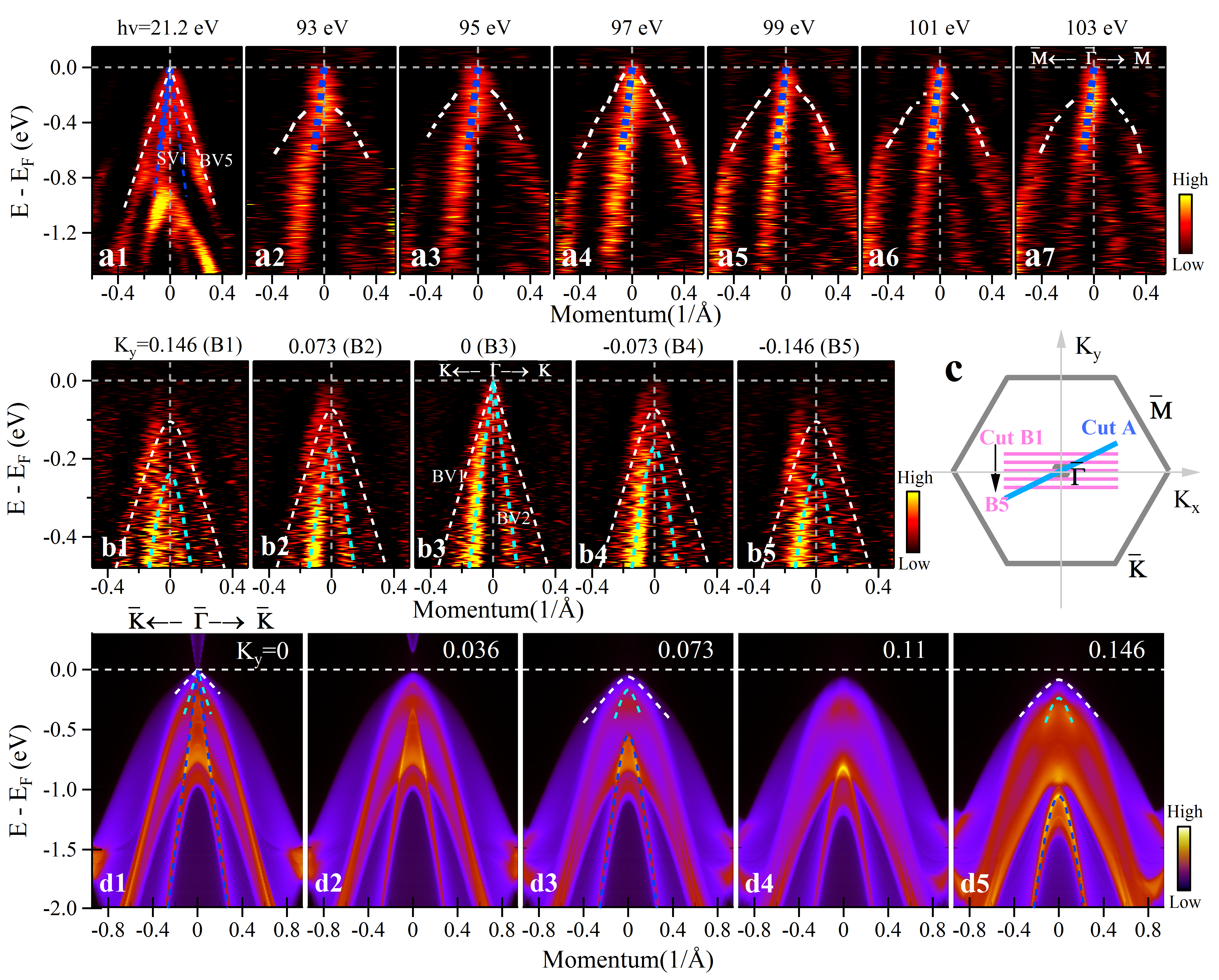}
\caption{\label{fig4} Band structure of Cd$_3$As$_2$ (112) surface measured at different K$_z$ and its band structure evolution in the K$_x$-K$_y$ plane.  (a). Band structure along $\Gamma$-M direction (cut A in (c)) measured using different photon energies at a temperature T = 45 K. Different photon energy corresponds to different K$_z$.  The inner band highlighted with blue dashed line shows weak K$_z$ dependence, while the outer band (marked as white dashed lines) shows an obvious change with K$_z$.   (b). Band structure along several momentum cuts along $\Gamma$-K direction (cuts B1 to B5 in (c)) around $\Gamma$ point in the K$_x$-K$_y$ plane measured using 21.2 eV photon energy.  The observed two sets of bands are fitted by hyperbolic function, shown as cyan dotted line for the inner band and white dahsed line for the outer band. The velocities used in the fitting functions are 1.4 eV$\cdot$ ${\AA}$ and 3.1 eV$\cdot$ ${\AA}$ for the outer band and inner band, respectively. (c). Location of the momentum cuts in the Brillouin zone. (d). Calculated band structures at different K$_y$ in the K$_x$-K$_y$ plane. Here, we use the same hyperbola functions for the calculated bands as those from fitting the measured bands in (b).
}
\end{figure*}

\newpage

\end{document}